\documentclass{elsart}
\usepackage{epsfig}
\usepackage{amssymb}
\usepackage{amsmath}

\newcommand{\affuni}[2]{Dipartimento di Fisica dell'Universit\`a 
  #1, #2, Italy.}
\newcommand{\affinfn}[2]{INFN Sezione di #1, #2, Italy.}

\newcommand{\dafne}     {DA$\Phi$NE }

\newcommand{\ep}{\mbox{$e^{+}$}}
\newcommand{\el}{\mbox{$e^{-}$}}
\newcommand{\pio}{\mbox{$\pi^{0}$}}
\newcommand{\pim}{\mbox{$\pi^{-}$}}
\newcommand{\pip}{\mbox{$\pi^{+}$}}
\newcommand{\ks}{\mbox{$K_{S}$}}
\newcommand{\kl}{\mbox{$K_{L}$}}

\begin{document}

\begin{frontmatter}

\title{\boldmath Search for a vector gauge boson in $\phi$ meson 
  decays with the KLOE detector}

\collab{The KLOE-2 Collaboration}
\author[Roma2,INFNRoma2]{F.~Archilli},
\author[Frascati]{D.~Babusci},
\author[Roma2,INFNRoma2]{D.~Badoni},
\author[Cracow]{I.~Balwierz},
\author[Frascati]{G.~Bencivenni},
\author[Roma1,INFNRoma1]{C.~Bini},
\author[Frascati]{C.~Bloise},
\author[INFNRoma1]{V.~Bocci},
\author[Frascati]{F.~Bossi},
\author[INFNRoma3]{P.~Branchini},
\author[Roma3,INFNRoma3]{A.~Budano},
\author[Moscow]{S.~A.~Bulychjev},
\author[Uppsala]{L.~Caldeira~Balkest\aa hl},
\author[Frascati]{P.~Campana},
\author[Frascati]{G.~Capon},
\author[Roma3,INFNRoma3]{F.~Ceradini},
\author[Frascati]{P.~Ciambrone},
\author[Frascati]{E.~Czerwi\'nski},
\author[Frascati]{E.~Dan\'e},
\author[Frascati]{E.~De~Lucia},
\author[INFNBari]{G.~De~Robertis},
\author[Roma1,INFNRoma1]{A.~De~Santis},
\author[Roma1,INFNRoma1]{G.~De~Zorzi},
\author[Roma1,INFNRoma1]{A.~Di~Domenico},
\author[Napoli,INFNNapoli]{C.~Di~Donato},
\author[Frascati]{D.~Domenici},
\author[Bari,INFNBari]{O.~Erriquez},
\author[Bari,INFNBari]{G.~Fanizzi},
\author[Frascati]{G.~Felici},
\author[Roma1,INFNRoma1]{S.~Fiore},
\author[Roma1,INFNRoma1]{P.~Franzini},
\author[Roma1,INFNRoma1]{P.~Gauzzi},
\author[Messina,INFNMessina]{G.~Giardina},
\author[Frascati]{S.~Giovannella\corauthref{cor}},
\ead{simona.giovannella@lnf.infn.it}
\corauth[cor]{Corresponding author.}
\author[Roma2,INFNRoma2]{F.~Gonnella},
\author[INFNRoma3]{E.~Graziani},
\author[Frascati]{F.~Happacher},
\author[Uppsala]{B.~H\"oistad},
\author[Frascati]{L.~Iafolla},
\author[Energetica,Frascati]{E.~Iarocci},
\author[Uppsala]{M.~Jacewicz},
\author[Uppsala]{T.~Johansson},
\author[Warsaw]{A.~Kowalewska},
\author[Moscow]{V.~Kulikov},
\author[Uppsala]{A.~Kupsc},
\author[Frascati,StonyBrook]{J.~Lee-Franzini},
\author[INFNBari]{F.~Loddo},
\author[Messina,INFNMessina]{G.~Mandaglio},
\author[Moscow]{M.~Martemianov},
\author[Frascati,Marconi]{M.~Martini},
\author[Roma2,INFNRoma2]{M.~Mascolo},
\author[Moscow]{M.~Matsyuk},
\author[Roma2,INFNRoma2]{R.~Messi},
\author[Frascati]{S.~Miscetti},
\author[Frascati]{G.~Morello},
\author[INFNRoma2]{D.~Moricciani},
\author[Cracow]{P.~Moskal},
\author[Roma3,INFNRoma3]{F.~Nguyen},
\author[INFNRoma3]{A.~Passeri},
\author[Energetica,Frascati]{V.~Patera},
\author[Roma3,INFNRoma3]{I.~Prado~Longhi},
\author[INFNBari]{A.~Ranieri},
\author[Uppsala]{C.~F.~Redmer},
\author[Frascati]{P.~Santangelo},
\author[Frascati]{I.~Sarra},
\author[Calabria,INFNCalabria]{M.~Schioppa},
\author[Frascati]{B.~Sciascia},
\author[Energetica,Frascati]{A.~Sciubba},
\author[Cracow]{M.~Silarski},
\author[Calabria,INFNCalabria]{S.~Stucci},
\author[Roma3,INFNRoma3]{C.~Taccini},
\author[INFNRoma3]{L.~Tortora},
\author[Frascati]{G.~Venanzoni},
\author[Frascati,CERN]{R.~Versaci},
\author[Warsaw]{W.~Wi\'slicki},
\author[Uppsala]{M.~Wolke},
\author[Cracow]{J.~Zdebik}.

\address[Bari]{\affuni{di Bari}{Bari}}
\address[INFNBari]{\affinfn{Bari}{Bari}}
\address[Calabria]{\affuni{della Calabria}{Cosenza}}
\address[INFNCalabria]{INFN Gruppo collegato di Cosenza, Cosenza, Italy.}
\address[Cracow]{Institute of Physics, Jagiellonian University, Cracow, Poland.}
\address[Frascati]{Laboratori Nazionali di Frascati dell'INFN, Frascati, Italy.}
\address[Messina]{\affuni{di Messina}{Messina}}
\address[INFNMessina]{\affinfn{Catania}{Catania}}
\address[Moscow]{Institute for Theoretical and Experimental Physics (ITEP), Moscow, Russia.}
\address[Napoli]{\affuni{''Federico II''}{Napoli}}
\address[INFNNapoli]{\affinfn{Napoli}{Napoli}}
\address[Energetica]{Dipartimento di Scienze di Base ed Applicate per l'Ingegneria dell'Universit\`a 
``Sapienza'', Roma, Italy.}
\address[Marconi]{Dipartimento di Scienze e Tecnologie applicate, Universit\`a "Guglielmo Marconi", Roma, Italy.}
\address[Roma1]{\affuni{''Sapienza''}{Roma}}
\address[INFNRoma1]{\affinfn{Roma}{Roma}}
\address[Roma2]{\affuni{``Tor Vergata''}{Roma}}
\address[INFNRoma2]{\affinfn{Roma Tor Vergata}{Roma}}
\address[Roma3]{\affuni{``Roma Tre''}{Roma}}
\address[INFNRoma3]{\affinfn{Roma Tre}{Roma}}
\address[StonyBrook]{Physics Department, State University of New 
York at Stony Brook, USA.}
\address[Uppsala]{Department of Physics and Astronomy, Uppsala University, Uppsala, Sweden.}
\address[Warsaw]{National Centre for Nuclear Research, Warsaw, Poland.}
\address[CERN]{Present Address: CERN, CH-1211 Geneva 23, Switzerland.}


\begin{abstract}

The existence of a light dark force mediator has been tested with 
the KLOE detector at DA$\Phi$NE.
This particle, called $U$, is searched for using the decay chain 
$\phi\to\eta\,U$, $\eta\to\pip\pim\pio$, $U\to\ep\el$. 
No evidence is found in 1.5 fb$^{-1}$ of data. The resulting exclusion 
plot covers the mass range $5<M_{U}<470$ MeV, setting an upper limit 
on the ratio between the $U$ boson coupling constant 
and the fine structure constant, $\alpha'/\alpha$, of 
$\leq 2\times 10^{-5}$ at 90\% C.L. for $50<M_{U}<420$ MeV.

\end{abstract}



\end{frontmatter}

\section{Introduction}
\label{Sec:Intro}

In recent years, several unexpected astrophysical observations have 
failed to find a common interpretation in terms of standard astrophysical 
or particle physics sources. A non-exhaustive list of these observations 
includes the 511 keV gamma-ray signal from the galactic center observed 
by the INTEGRAL satellite \cite{Jean:2003ci}, the excess in the cosmic 
ray positrons reported by PAMELA~\cite{Adriani:2008zr}, the total electron 
and positron flux measured by ATIC~\cite{Chang:2008zzr}, 
Fermi~\cite{Abdo:2009zk}, and HESS 
\cite{Collaboration:2008aaa,Aharonian:2009ah}, the annual modulation 
of the DAMA/LIBRA signal~\cite{Bernabei:2005hj,Bernabei:2008yi} and
the low energy spectrum of nuclear recoil candidate events observed
 by CoGeNT \cite{COGENT}.

Although there are alternative explanations for some of these anomalies, 
they could be all explained with the existence of a dark matter weakly 
interacting massive particle, WIMP, belonging to a secluded gauge sector 
under which the Standard Model (SM) particles are uncharged
\cite{Pospelov:2007mp,ArkaniHamed:2008qn,Alves:2009nf,Pospelov:2008jd,%
Hisano:2003ec,Cirelli:2008pk,MarchRussell:2008yu,Cholis:2008wq,%
Cholis:2008qq,ArkaniHamed:2008qp}.
An abelian gauge field, the $U$ boson with mass near the GeV scale, 
couples the secluded sector to the SM through its kinetic mixing with 
the SM hyper-charge gauge field. The kinetic mixing parameter, $\epsilon$, 
is expected to be of the order 10$^{-4}$--10$^{-2}$
\cite{ArkaniHamed:2008qn,Essig:2009nc}, 
so that observable effects can be induced in 
$\mathcal{O}(\mbox{GeV}$)--energy $e^+e^-$ colliders 
\cite{Essig:2009nc,Batell:2009yf,Reece:2009un,
Borodatchenkova:2005ct,Yin:2009mc} and fixed target experiments 
\cite{Bjorken:2009mm,Batell:2009di,Essig:2010xa,Freytsis:2009bh}. 
The possible existence of a new light boson gauging a new symmetry with 
a small coupling was in fact already introduced on general grounds in 
\cite{Fayet1}, and rediscussed in  models postulating also the existence 
of light spin 0 or 1/2 dark matter particles \cite{Fayet2,Fayet3}. This 
boson can have both vector and axial-vector couplings to quark and leptons, 
however axial couplings are strongly constrained by data, leaving room 
to vector couplings only.

The $U$ boson can be produced at \ep\el\ colliders via different 
processes: $e^+e^-\to U\gamma$, $e^+e^-\to Uh'$ ($h'$-strahlung), 
where $h'$ is a higgs-like particle responsible for the breaking 
of the hidden symmetry, and $V\to P\gamma$ decays, where $V$ and $P$ 
are vector and pseudoscalar mesons, respectively. 
In this work, we study the process $\phi \rightarrow \eta\,U$, using
a sample of $\phi$ mesons produced resonantly at the \dafne\ collider.
The $U$ boson can be observed by its decay into a lepton pair, while the 
$\eta$ can be tagged by one of its main decays. 
An irreducible background due to the Dalitz decay of the $\phi$ meson, 
$\phi\to\eta\,\ell^+\ell^-$, is present. This decay has been studied by 
the SND and CMD-2 experiments, which measured a branching fraction of
${\mbox BR}(\phi\to\eta\,e^+e^-) = ( 1.19 \pm 0.19 \pm 0.07 )\times 10^{-4}$ 
and 
${\mbox BR}(\phi\to\eta\,e^+e^-) = ( 1.14 \pm 0.10 \pm 0.06 )\times 10^{-4}$,
respectively \cite{phietaeeSND,phietaeeCMD2}. This corresponds to a cross 
section of $\sigma(\phi\to\eta\,\ell^+\ell^-)\sim 0.7$ nb, with a di-lepton
mass range $M_{\ell\ell}<470$ MeV. For the signal, the expected cross 
section is expressed by \cite{Reece:2009un}:
\begin{equation}
  \sigma(\phi\to\eta\,U) = \epsilon^2 \, |F_{\phi\eta}(m_U^2)|^2 \,
  \frac{\lambda^{3/2}(m_\phi^2,m_\eta^2,m_U^2)}
       {\lambda^{3/2}(m_\phi^2,m_\eta^2,0)}
  \, \sigma(\phi\to\eta\gamma) \,, 
  \label{Eq:Reece}
\end{equation}
where $F_{\phi\eta}(m_U^2)$ is the $\phi\eta\gamma^*$ transition form 
factor evaluated at the $U$ mass while the following term represents 
the ratio of the kinematic functions of the involved decays.\!%
\footnote{$\lambda(m_1^2,m_2^2,m_3^2) = [1 + m_3^2/(m_1^2-m_2^2)]^2 - 
4 m_1^2 m_3^2/(m_1^2-m_2^2)^2$.}
Using $\epsilon = 10^{-3}$ and $|F_{\phi\eta}(m_U^2)|^2 =1$, a cross 
section $\sigma(\phi\to\eta\,U)\sim 40\ {\rm fb}$ is obtained. Despite 
the small ratio between the overall cross section of $\phi\to\eta\,U$ 
and $\phi\to\eta\,\ell^+\ell^-$, their different di-lepton invariant 
mass distributions allow to test the $\epsilon$ parameter down to 
$10^{-3}$ with the KLOE data set.

The best $U$ decay channel to search for the $\phi\to\eta\,U$ process 
at KLOE is in $e^+e^-$, since a wider range of $U$ boson masses can be 
tested and $e^\pm$ are easily identified using a time-of-flight (ToF) 
technique. The $\eta$ can be tagged by the three-pion or two-photon final 
state, which represent $\sim85\%$ of the total decay rate. We have used 
the $\eta\to\pi^+\pi^-\pi^0$ decay channel, which provides a clean final 
state with four charged particles and two photons.

\section{The KLOE detector}

The KLOE experiment operated from 2000 to 2006 at \dafne, the Frascati 
$\phi$-factory. DA$\Phi$NE is an $e^+e^-$ collider running at a 
center-of-mass energy of $\sim 1020$~MeV, the mass of the $\phi$ meson. 
Equal energy positron and electron beams collide at an angle of $\pi$-25 
mrad, producing $\phi$ mesons nearly at rest.
The detector consists of a large cylindrical Drift Chamber (DC),
surrounded by a lead-scintillating fiber electromagnetic calorimeter
(EMC). A superconducting coil around the EMC provides a 0.52~T field.
The beam pipe at the interaction region is spherical in shape  with 
10 cm radius, it is made of a Beryllium-Aluminum alloy of 0.5 mm
thickness.
Low beta quadrupoles are located at about $\pm$50 cm distance from the
interaction region.
The drift chamber~\cite{DCH}, 4~m in diameter and 3.3~m long, has 12,582
all-stereo tungsten sense wires and 37,746 aluminum field wires. 
The chamber shell is made of carbon fiber-epoxy composite with an
internal wall of $\sim 1$ mm thickness, the gas used is a 90\% helium, 
10\% isobutane mixture. 
The spatial resolutions are $\sigma_{xy} \sim 150\ \mu$m and 
$\sigma_z \sim$~2 mm.
The momentum resolution is $\sigma(p_{\perp})/p_{\perp}\approx 0.4\%$.
Vertexes are reconstructed with a spatial resolution of $\sim$ 3~mm.
The calorimeter~\cite{EMC} is divided into a barrel and two endcaps, for a
total of 88 modules, and covers 98\% of the solid angle. 
The modules are read out at both ends by photomultipliers, both in
amplitude and time. 
The readout granularity is $\sim$\,(4.4 $\times$ 4.4)~cm$^2$, for a total
of 2440 cells arranged in five layers. 
The energy deposits are obtained from the signal amplitude while the
arrival times and the particles positions are obtained from the time
differences. 
Cells close in time and space are grouped into energy clusters. 
The cluster energy $E$ is the sum of the cell energies.
The cluster time $T$ and position $\vec{R}$ are energy-weighted averages. 
Energy and time resolutions are $\sigma_E/E = 5.7\%/\sqrt{E\ {\rm(GeV)}}$ 
and  $\sigma_t = 57\ {\rm ps}/\sqrt{E\ {\rm(GeV)}} \oplus100\ {\rm ps}$, 
respectively.
The trigger \cite{TRG} uses both calorimeter and chamber information.
In this analysis the events are selected by the calorimeter trigger,
requiring two energy deposits with $E>50$ MeV for the barrel and $E>150$
MeV for the endcaps. A cosmic veto rejects events with at least two 
energy deposits above 30 MeV in the outermost calorimeter layer. 
Data are then analyzed by an event classification filter \cite{NIMOffline},
which selects and streams various categories of events in different
output files.

\section{\boldmath Event selection}
\label{Sec:DataSample}

The analysis of the decay chain $\phi\to\eta\,U$, $\eta\to\pip\pim\pio$, 
$U\to e^+ e^-$, has been performed on a data sample of 1.5 fb$^{-1}$,
corresponding approximately to $5\times 10^9$ produced $\phi$ mesons.
The Monte Carlo (MC) simulation of the irreducible background 
$\phi\to\eta\,\ep\el$, $\eta\to\pip\pim\pio$ has been produced with 
$d\Gamma(\phi\to\eta\,\ep\el)/dm_{ee}$ weighted according to Vector 
Meson Dominance model \cite{Landsberg85}, using the form factor 
parametrization from the SND experiment \cite{phietaeeSND}.
The MC simulation for the $\phi\to\eta\, U$ decay has been developed 
according to \cite{Reece:2009un}, with a flat distribution in $M_{ee}$.
All MC productions, including all other $\phi$ decays, take into account 
changes in \dafne\ operation and background conditions on a run-by-run 
basis. Data-MC corrections for cluster energies and tracking efficiency, 
evaluated with radiative Bhabha events and $\phi\to\rho\pi$ samples,
respectively, have been applied.

\begin{figure}[!t]
  \begin{center}
    \epsfig{file=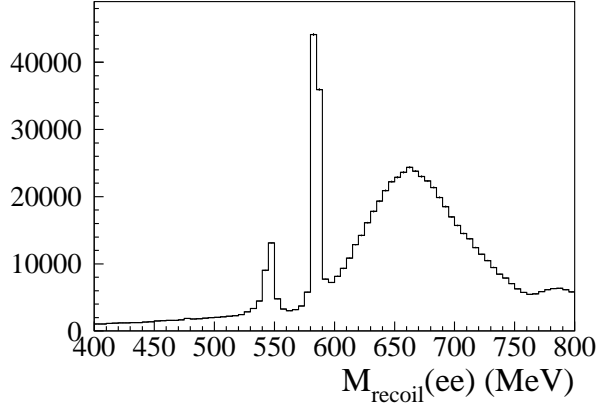,width=0.6\textwidth}
  \end{center}
  \caption{Recoiling mass against the $e^+e^-$ pair for the data sample 
    after preselection cuts. The $\phi\to\eta\,e^+e^-$ signal is clearly 
    visible in the peak corresponding to $\eta$ mass. The second peak 
    at $\sim 590$ MeV is due to $\phi\to\ks\kl$, $K_S\to\pi^+\pi^-$ 
    events with wrong mass assignment.}
  \label{Fig:Mmiss}
\end{figure}

As a first step of the analysis, a preselection is performed requiring:
\begin{enumerate}
\item two positive and two negative tracks with point of closest 
  approach to the beam line inside a cylinder around the interaction 
  point (IP), with transverse radius $R_{\rm FV}=4$ cm and length 
  $Z_{\rm FV}=20$ cm;
\item two photon candidates {\it i.e.} two energy clusters with
  $E > 7$ MeV not associated to any track, in an angular acceptance 
  $|\cos\theta_\gamma|<0.92$ and in the expected time window for a 
  photon
  ($|T_\gamma-R_\gamma/c|<{\rm MIN}(5\sigma_T,2\,{\rm ns})$); 
\item best $\pip\pim\gamma\gamma$ match to the $\eta$ mass in the pion 
  hypothesis to assign $\pi^\pm$ tracks;\!%
  \footnote{The invariant mass of $\pip\pim\gamma\gamma$ for each 
    positive/negative track pair, $M_{\rm test}$, is evaluated in the 
    hypothesis that the two tracks belong to charged pions. The track 
    pair with the smaller $|M_{\rm test}-M_\eta|$ value is assigned to 
    \pip\ and \pim.}
  the other two tracks are then assigned to $e^\pm$; 
\item loose cuts of about $\pm\, 4\,\sigma$'s on $\eta$ and $\pi^0$ 
  invariant masses ($495 < M_{\pi^+\pi^-\gamma\gamma} < 600$ MeV, 
  $70 < M_{\gamma\gamma} < 200$ MeV).
\end{enumerate}
After this selection, a clear peak corresponding to $\phi\to\eta\,\ep\el$ 
events is observed in the distribution of the recoil mass to the $e^+e^-$ 
pair, $M_{\rm recoil}(ee)$, as shown in Fig.~\ref{Fig:Mmiss}. 
The second peak at $\sim 590$ MeV is due to $K_S\to\pi^+\pi^-$ events 
with a wrong mass assignment. Events in the $535 < M_{\rm recoil}(ee) < 560$ 
MeV window are retained for further analysis.

\begin{figure}[!t]
  \begin{center}
    \epsfig{file=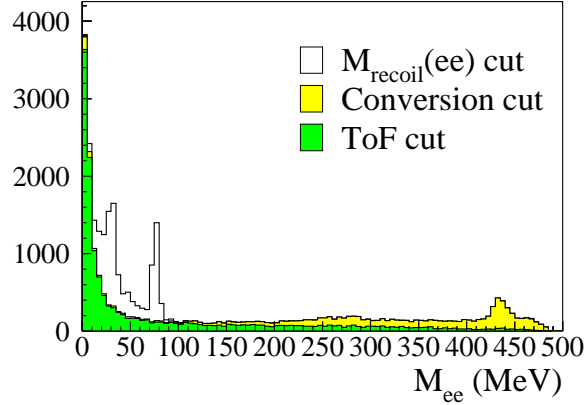,width=0.6\textwidth}
  \end{center}
  \caption{$M_{ee}$ distribution for data after different analysis 
    cuts.}
  \label{Fig:BckgReduction}
\end{figure}

\begin{figure}[!t]
  \begin{center}
    \epsfig{file=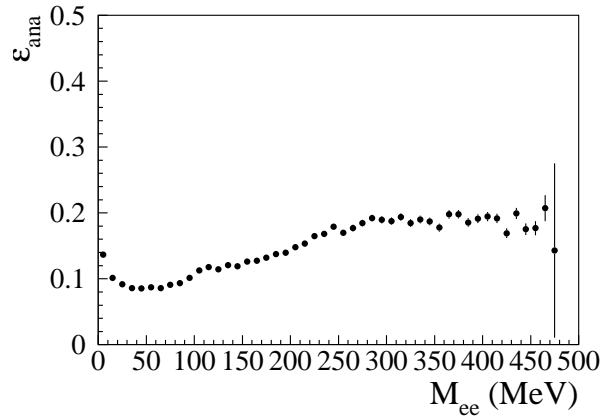,width=0.6\textwidth}
  \end{center}
  \caption{Analysis efficiency as a function of \ep\el\ invariant mass.}
  \label{Fig:Effi}
\end{figure}

A residual background contamination, due to $\phi\to\eta\gamma$ events
with photon conversion on beam pipe (BP) or drift chamber walls (DCW), 
is rejected by tracing back the tracks of the two $e^+$, $e^-$ candidates 
and reconstructing their invariant mass ($M_{ee}$) and distance ($D_{ee}$)
at the BP/DCW surfaces. As both quantities are small in case of photon 
conversions, $\phi\to\eta\gamma$ background is removed by rejecting 
events with: 
$M_{ee}(BP) < 10$ MeV and $D_{ee}(BP) < 2$ cm, 
$M_{ee}(DCW)< 80$ MeV and $D_{ee}(DCW)< 10$ cm.
A further relevant background, originated by $\phi\to K \bar{K}$
decays surviving analysis cuts, has more than two charged
pions in the final state and is suppressed using time-of-flight (ToF) 
to the calorimeter. When an energy cluster is connected to a track, the 
arrival time to the calorimeter is evaluated both using the calorimeter 
timing ($T_{\rm cluster}$) and the track trajectory 
($T_{\rm track}=L_{\rm track}/\beta c$). 
The $\Delta T = T_{\rm track}-T_{\rm cluster}$ variable is then evaluated 
in both electron ($\Delta T_e$) and pion ($\Delta T_\pi$) hypotheses. 
Events with an $e^+$, $e^-$ candidate outside a 3\,$\sigma$'s window on 
the $\Delta T_e$ variables are rejected. 
In Fig.~\ref{Fig:BckgReduction}, the $M_{ee}$ distribution evaluated 
at different steps of the analysis is shown. 
The peaks at $\sim 30$ MeV and $\sim 80$ MeV are due to photon conversions 
on BP and DCW, respectively.
The ToF cut reduces the tail at high $M_{ee}$ values while the 
conversion cut removes events in the low invariant mass region. 
The analysis efficiency, defined as the ratio between the number of
events surviving analysis cuts and that of all generated events, 
is shown in Fig.~\ref{Fig:Effi} as a function of $M_{ee}$,
ranging between 10\% and 20\%. The main contribution to the loss 
of efficiency is due to preselection cuts, being
$\varepsilon_{\rm presel} = (24.73 \pm 0.04)\%$.

\begin{figure}[!t]
  \begin{center}
    \epsfig{file=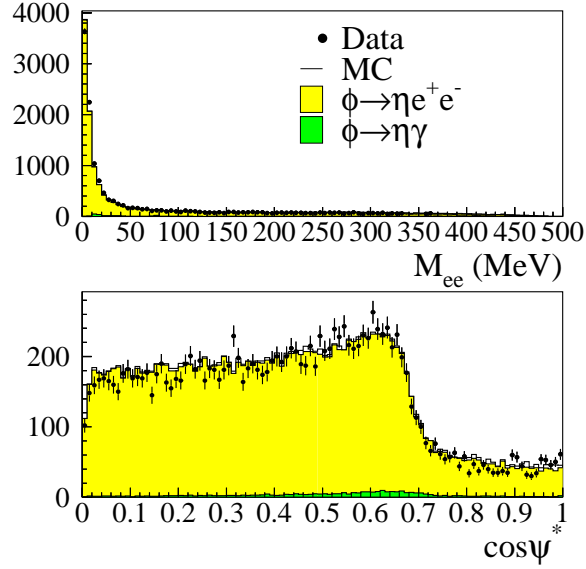,width=0.6\textwidth}
  \end{center}
  \caption{Invariant mass of the $e^+ e^-$ pair (top) and $\cos\psi^*$ 
    distribution (bottom) for $\phi\to\eta\,\ep\el$, $\eta\to\pip\pim\pio$ 
    events. Dots are data, the black solid line is the sum of all MC 
    expectations while signal and residual background contamination from 
    $\phi\to\eta\gamma$ are shown in colors.}
  \label{Fig:AllCuts}
\end{figure}

In Fig.~\ref{Fig:AllCuts} the comparison between data and Monte Carlo
events for $M_{ee}$ and $\cos\psi^*$ distributions is shown. Here 
$\psi^*$ is the angle between the $\eta$ and the $e^+$ in the $e^+e^-$ 
rest frame. About 14,000 $\phi\to\eta\,e^+e^-$, $\eta\to\pip\pim\pio$ 
candidates are present in the analyzed data set, with a negligible 
residual background contamination. 

\section{\boldmath Upper limit evaluation}

As an accurate description of the background is crucial for the search 
of the $U$ boson, its shape is extracted directly from our data. A fit 
is performed to the $M_{ee}$ distribution, after applying a bin-by-bin 
subtraction of the $\phi\to\eta\gamma$ background and efficiency 
correction. The parametrization of the fitting function has been taken 
from Ref.~\cite{Landsberg85}:
\begin{eqnarray}
  \frac{d\Gamma(\phi\to\eta\,e^+e^-)}{dq^2}
  = \ \ \ \ \ \ \ \ \ \ \ \ \ \ \ \ \ \ \ \ \ \ \ \ 
  \ \ \ \ \ \ \ \ \ \ \ \ \ \ \ \ \ \ \ \ \ \ \ \ \ 
  \ \ \ \ \ \ \ \ \nonumber \\
  \frac{\alpha}{3\pi}\frac{|F_{\phi\eta}(q^2)|^2}{q^2} 
  \sqrt{1-\frac{4m^2}{q^2}} \left(1+\frac{2m^2}{q^2}\right) 
  \lambda^{3/2}(m_\phi^2,m_\eta^2,m_U^2)
\end{eqnarray}
with $q=M_{ee}$ and the transition form factor described by

\begin{equation}
  F_{\phi\eta}(q^2) = \frac{1}{1-q^2/\Lambda^2} \, .
 \label{Eq:FF}
\end{equation}

\begin{figure}[!t]
  \begin{center}
    \epsfig{file=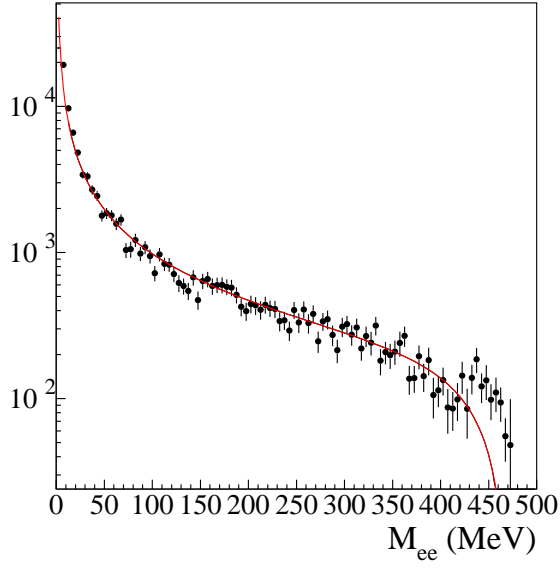,width=0.6\textwidth}
  \end{center}
  \caption{Fit to the corrected $M_{ee}$ spectrum for the Dalitz decays
    $\phi\to\eta\,e^+e^-$.}
  \label{Fig:Fit}
\end{figure}

Free parameters of the fit are $\Lambda$ and an overall normalization 
factor. A good description of the $M_{ee}$ shape 
is obtained except at the high end of the 
spectrum (see Fig.~\ref{Fig:Fit}), 
where a residual background contamination from multi-pion events
is still present.

\begin{figure}[!t]
  \begin{center}
    \epsfig{file=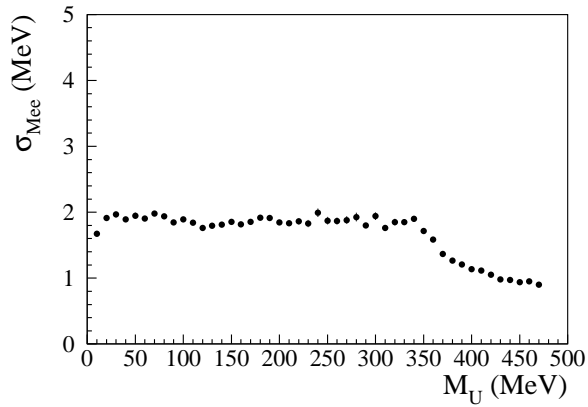,width=0.6\textwidth}
  \end{center}
  \caption{Resolution on $M_{ee}$ as a function of the $U$ boson 
    invariant mass for $\phi\to\eta\,U$ MC events.}
  \label{Fig:ResoMee}
\end{figure}

As mentioned in Sec.~\ref{Sec:DataSample}, the $\phi\to\eta\,U$ MC 
signal has been produced according to Ref.~\cite{Reece:2009un}, with 
a flat distribution of the $U$ boson invariant mass, $M_U$.
This sample has been used to evaluate the resolution on the \ep\el\
invariant mass as a function of $M_U$, applying a Gaussian fit to the 
$M_{ee}-M_{U}$ distributions. Results are reported in Fig.~\ref{Fig:ResoMee}. 
The resolution is $\sim 2$ MeV for $M_U < 350$ MeV and then improves
to 1 MeV for higher values.
The upper limit on $\phi\to\eta\,U$ signal as a function of $M_{U}$ 
is then obtained in the following way:
\begin{itemize}
\item[(a)] MC events are divided in sub-samples of 1 MeV width in 
  the range $5 < M_{U} < 470$ MeV;
\item[(b)] for each $M_{U}$ sub-sample, the average value of the 
  $\phi\to\eta\,\ep\el$ background, $b(M_{ee})$, is obtained by 
  fitting the reconstructed $M_{ee}$ spectrum with 5 MeV binning, 
  removing five bins centered at $M_{U}$;
\item[(c)] for each fit, the maximum variation of $b(M_{ee})$ events,
  $\Delta b(M_{ee})$, is obtained changing by $\pm 1\,\sigma$ the fit
  parameters;
\item[(d)] for each $M_{U}$ value, the signal hypothesis is tested 
  comparing observed data, $b(M_{ee})$ and MC signal in the five
  reconstructed bins excluded in (b). The exclusion plot is obtained
  applying the CLs method \cite{CLS}. A Gaussian spread of width
  $\Delta b(M_{ee})$ on the background distribution is applied while
  evaluating CLs.
\end{itemize}

\begin{figure}[!t]
  \begin{center}
    \epsfig{file=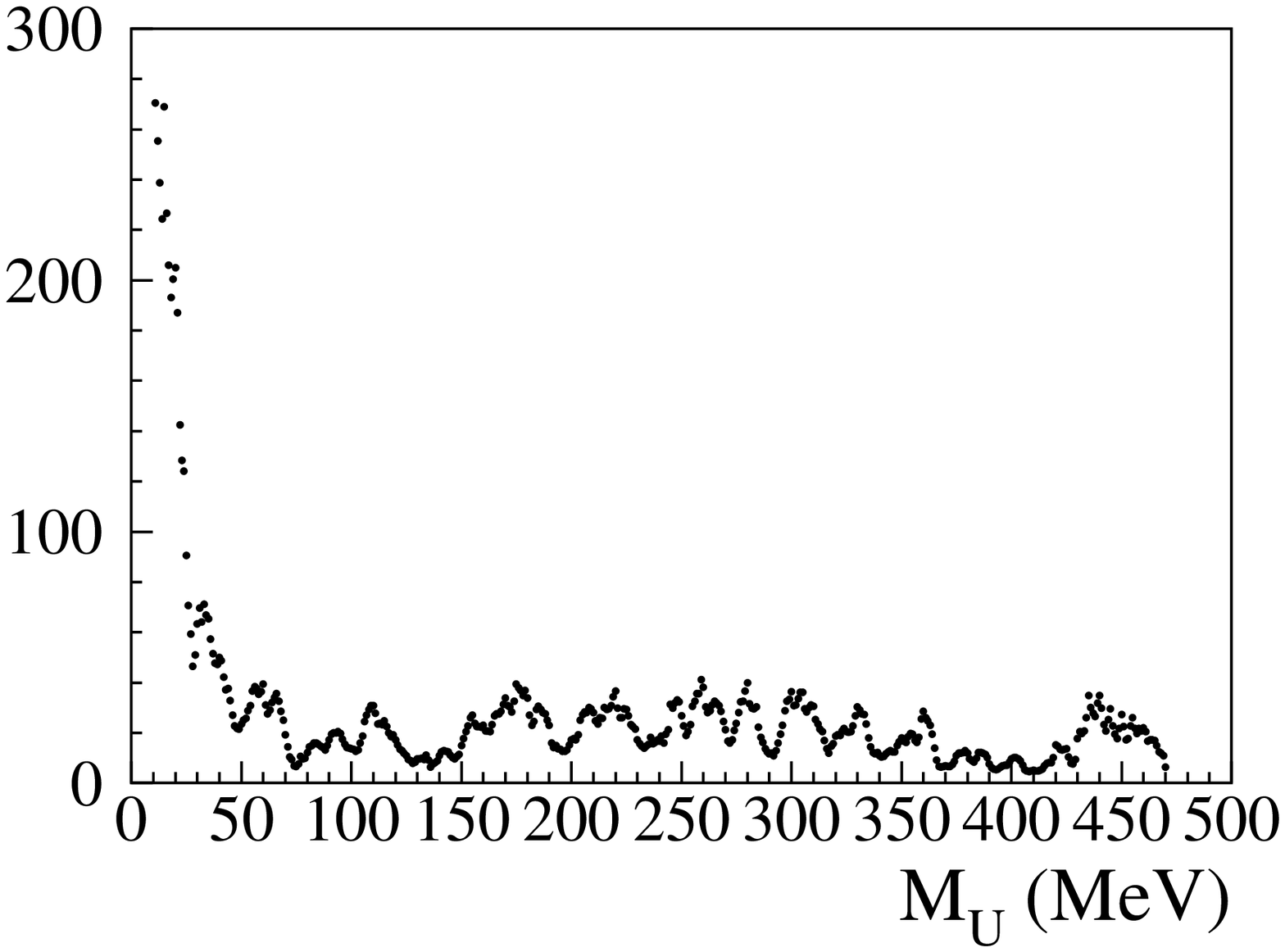,width=0.6\textwidth} 
  \end{center}
  \caption{Upper limit at 90\% C.L. on the number of events for the 
    decay chain $\phi\to\eta\,U$, $\eta\to\pip\pim\pio$, $U\to\ep\el$.}
  \label{Fig:UL_nev}
\end{figure}

\begin{figure}[!t]
  \begin{center}
    \epsfig{file=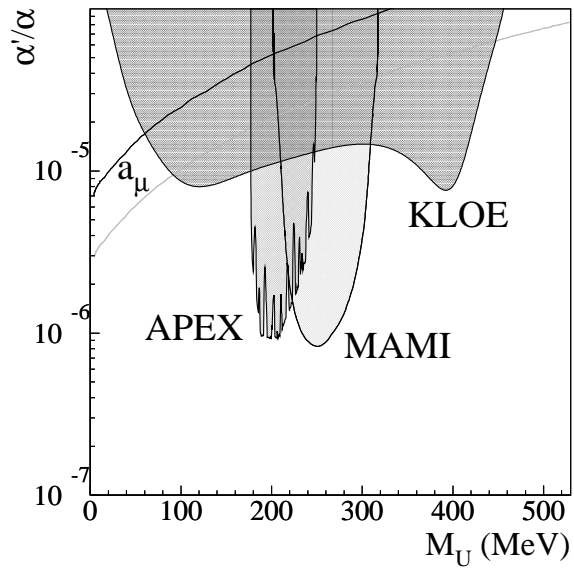,width=0.6\textwidth} 
  \end{center}
  \caption{Exclusion plot at 90\% C.L. for the parameter 
    $\alpha'/\alpha=\epsilon^2$, compared with existing limits in
    our region of interest.}
  \label{Fig:ULcurves}
\end{figure}

In Fig.~\ref{Fig:UL_nev} the exclusion plot at 90\% C.L.\ on the number 
of events for the decay chain $\phi\to\eta\,U$, $\eta\to\pip\pim\pio$, 
$U\to\ep\el$, is shown. 
Using Eq.~(\ref{Eq:Reece}) and taking into account the analysis 
efficiency this result is then reported in terms of the parameter 
$\alpha'/\alpha=\epsilon^2$, where $\alpha'$ is the coupling of the 
$U$ boson to electrons and $\alpha$ is the fine structure constant.
The opening of the $U\to\mu^+\mu^-$ threshold, in the hypothesis that 
the $U$ boson decays only to lepton pairs and assuming 
equal coupling to  \ep\el\ and $\mu^+\mu^-$,
has been included.
In Fig.~\ref{Fig:ULcurves} the smoothed exclusion plot at 90\% C.L.\ 
on $\alpha'/\alpha$ is compared with existing limits from the muon
anomalous magnetic moment $a_\mu$ \cite{amu} and from recent measurements 
of the MAMI/A1 \cite{UL_MAMI} and APEX \cite{UL_APEX} experiments.
The gray line is where the $U$ boson parameters should lay to account 
for the observed discrepancy between measured and calculated $a_\mu$
values.
Our result greatly improves existing limits in a wide mass range, 
resulting in an upper limit on the $\alpha'/\alpha$ parameter of 
$\leq 2\times 10^{-5}$ @ 90\% C.L.\ for $50<M_{U}<420$ MeV, thus 
covering part of the expected $\epsilon$ range (see Sec.~\ref{Sec:Intro}).
We exclude that the existing $a_\mu$ discrepancy is due to a $U$ boson 
with mass ranging between 90 and 450 MeV. 

\section*{Acknowledgments}

We warmly thank our former KLOE colleagues for the access to the data 
collected during the KLOE data taking campaign.
We thank the DA$\Phi$NE team for their efforts in maintaining low 
background running conditions and their collaboration during all 
data taking. 
We want to thank our technical staff: 
G.~F.~Fortugno and F.~Sborzacchi for their dedication in ensuring
efficient operation of the KLOE computing facilities;
M.~Anelli for his continuous attention to the gas system and detector
safety;
A.~Balla, M.~Gatta, G.~Corradi and G.~Papalino for electronics
maintenance;
M.~Santoni, G.~Paoluzzi and R.~Rosellini for general detector support;
C.~Piscitelli for his help during major maintenance periods.
This work was supported in part
by the EU Integrated Infrastructure Initiative HadronPhysics Project
under contract number RII3-CT-2004-506078;
by the European Commission under the 7th Framework Programme through
the 'Research Infrastructures' action of the 'Capacities' Programme,
Call: FP7-INFRASTRUCTURES-2008-1, Grant Agreement N. 227431;
by the Polish Ministery of Science and Higher Education through the
Grant No. 0469/B/H03/2009/37.


\end{document}